\documentstyle[11pt]{article}

\begin{document}

\def\mxth{\mathsurround=0pt }
\def\xversim#1#2{\lower2.pt\vbox{\baselineskip0pt \lineskip-.5pt
\ialign{$\mxth#1\hfil##\hfil$\crcr#2\crcr\sim\crcr}}}
\def\simgr{\mathrel{\mathpalette\xversim >}}
\def\simle{\mathrel{\mathpalette\xversim <}}

\title
{AN UPPER BOUND ON THE HIGGS BOSON MASS
        \\FROM YUKAWA UNIFICATION\\
AND A COMMENT ON VACUUM STABILITY CONSTRAINTS}

\author{Nir Polonsky
\thanks{Talk presented at SUSY-94,
International Workshop  on Supersymmetry and Unification
of Fundamental Interactions,
Ann Arbor, Michigan, May 14 -17, 1994. Pennsylvania Report No. UPR-0595T.}
\\
{\it Department of Physics, University of Pennsylvania,
Philadelphia, Pennsylvania,  19104}
}
\date{May 1994}
\maketitle

\begin{abstract}
Only small regions in the $m_{t} - \tan\beta$ plane are allowed
when considering simultaneously (assuming the MSSM)
coupling constant unification and  (minimal) GUT
relations among Yukawa couplings (i.e., $h_{b} = h_{\tau}$
at the unification point). In particular, if $m_{t} \simle 175$ GeV
we find that only $1 \simle \tan\beta \simle 1.5$
or $\tan\beta \simgr 40 \pm 10$ is allowed. The former implies that
the light Higgs boson is $\simle 110$ GeV and, in  principle, visible to LEPII.
The prediction for the Higgs boson mass in the $\tan\beta \approx 1$
scenario is discussed and uncertainties related to  ($i$)
vacuum stability constraints, ($ii$) different methods for calculating
the Higgs boson mass, ($iii$) two-loop calculations and  ($iv$)
GUT corrections are briefly reviewed. It is shown that large left-right
mixing between the $t$-scalars can significantly enhance the Higgs boson mass.
That and an ambiguity in the size of the two-loop correction lead to our
conservative  upper bound of 110 GeV. Vacuum stability considerations
constrain the $t$-scalar mixing and slightly diminish the upper bound
(depending on the value of $m_{t}$). Improved two-loop
calculations are also expected to strengthen the bound.
\end{abstract}

Realizing the minimal supersymmetric
extention of the standard model (MSSM)
within a grand-unified theory (GUT)
and assuming minimal matter content,
often imply specific relations between
the different Yukawa couplings. When considering the
simplest example, i.e.,
``bottom-tau unification'' \cite{hbhtau} $h_{b} = h_{\tau}$,
we find \cite{us} (in agreement with others \cite{others}) that
the predicted ranges for
$\alpha_{s}(M_{Z})$
and for
$\tan\beta = \nu_{up}/\nu_{down}$
are strongly correlated with the
weak angle
$s^{2}(M_{Z}) \approx 0.2324$\footnote{For $m_{t}^{pole} \approx 143$ GeV.}
and with $m_{t}$, and
are strongly constrained\footnote{
$\alpha_{s}$, $\tan\beta$ and $m_{t}$ are the
strong coupling,
the ratio of the two Higgs doublet expectation values and
$t$-quark mass, respectively.}.
Such relations provide a consistency test of simple GUT models,
as well as a means to probe the GUT structure.
Furthermore, they can be used to distinguish such models
from generic string models that typically do not constrain
the Yukawa couplings in this manner.
Thus, exploration of all implications of the
bottom-tau unification
(and similar relations)
is well motivated. Below, we study
the implications for the mass of the light MSSM (CP even) Higgs boson
$m_{h^{0}}$.
We will motivate the exploration of the
$\tan\beta \rightarrow 1$ limit
and show that $m_{h^{0}}$
in this region is heavier
than naive expectations, but most probably not too
heavy to be seen in LEPII \cite{penn594}. Exploring that region,
where the SM minimum $\propto \cos^{2}2\beta$
is shallow, requires
us to pay special attention to vacuum stability constraints
and we will briefly comment on that issue.

%2222222

The most impressive evidence for TeV scale supersymmetry
is the well publicized coupling constant unification \cite{ccunification}.
However, it is important to note that
[up to matching functions induced by GUT-scale physics, which
dominate the $0.008$ theoretical error bar in (\ref{als})]
unification requires $\alpha_{s}(M_{Z}) \simgr 0.12$, i.e.,
\begin{equation}
\alpha_{s}(M_{Z}) = 0.125 \pm 0.001 \pm 0.008 + H_{\alpha_{s}}
+ 3.2 \times 10^{-7}{\mbox{GeV}}^{-2}[(m_{t}^{pole})^{2}
- (143\,{\mbox{GeV}})^{2}].
\label{als}
\end{equation}
The heavier the $t$-quark,
the more the extracted value for the weak-angle is diminished
and the predicted value for $\alpha_{s}$ is increased.
Two-loop contributions from Yukawa interactions $H_{\alpha_{s}}$
can diminish
$\alpha_{s}$ by as much as 0.003 ($\sim 2.5\%$) if the Yukawa coupling
is of order unity
(i.e., the function $H_{\alpha_{s}}\simgr -0.003$).
The above observations affect the region in parameter space
which is consistent with the observational range for the $b$-quark
mass and with  bottom-tau unification. In particular, for
$m_{t}^{pole} \simle 200$ GeV only two regions are allowed:
($i$) $\tan\beta \approx 1$  (see Fig. 1) and ($ii$) large $\tan\beta$.
For example, if $m_{t}^{pole} \approx 174$ GeV \cite{cdf}
then $\tan\beta \approx 1.5$ or $\tan\beta \approx 40 \pm 10$.
Hereafter, I will discuss region ($i$) only.
An important implication of Yukawa unification,
choosing the $\tan\beta$ near unity region, is that the
SM-like Higgs boson mass nearly vanishes at tree-level. Thus, a careful
examination of the loop corrections is required.

There are important theortical uncertainties which
constrain the predictive power of the model \cite{us}. These are taken
into account (assuming no conspiracies),e.g., in (\ref{als}) and Fig. 1,
but will not be discussed here.
The following points, however, should be stressed:

\begin{itemize}
\item
$\alpha_{s}$ dependence:
We use the predicted value of $\alpha_{s}(M_{Z})[s^{2}, m_{t},\, ...]$,
e.g.,
$\alpha_{s}(M_{Z}) \approx 0.12$ for $m_{t}^{pole}\approx 100$ GeV
and
$\alpha_{s}(M_{Z}) \approx 0.13$ for $m_{t}^{pole}\approx 180$ GeV.
Using instead a fixed value $\alpha_{s}^{0}(M_{Z})$
for the whole $m_{t}$ range
implies some hidden assumptions on GUT-scale corrections.
The lower the value of $\alpha_{s}$ (and the heavier the $t$-quark)
the larger and less likely are these corrections.
\item
Finite-loop corrections to $m_{b}(M_{SUSY})$:
The relevance of those corrections was
pointed out \cite{delmb} recently.
However, while being extremely important for large $\tan\beta$
(maybe even to the point of removing any predictive power),
they are negligible in region ($i$), as can be seen in Fig. 2.
The details of the low-energy spectrum are nearly
irrelevant in determining the size of that region.

\end{itemize}

%3333333333333

Different aspects of region ($i$) were discussed by various authors.
The possibility that the Higgs boson mass is induced
only at the loop-level was discussed
by Diaz and Haber \cite{haberdiaz}. The consistency
of that scenario with
bottom-tau unification recently led to greater attention
to that region of parameter space \cite{penn594,others2,dreesetal}.
The consistency with
Yukawa unification is easily understood:
{}From perturbativity one has a $m_{t}$-dependent
lower bound on $\tan\beta \simgr 1$. Thus, we are in a region of large
top-Yukawa coupling $h_{t} \approx 1$. Slightly decreasing $\tan\beta$ (i.e.,
decreasing $\sin\beta$ or increasing $h_{t}$)
at $M_{Z}$ would lead to
divergences  (i.e., $h_{t}$ ``hits'' its Landau pole)
at higher  scales \cite{hill}:
$h_{t}$ has a quasi-fixed point at $\tan\beta \approx 1$.
This $\underline{\mbox{balances
too large $\alpha_{s}$ corrections to the $h_{b}/h_{\tau}$ ratio}}$,
can explain the heavy $t$-quark,
and makes region ($i$) relatively insensitive to theoretical
uncertainties.

For $\tan\beta = 1$ there is a custodial symmetry
$SU(2)_{L}\times SU(2)_{R} \rightarrow SU(2)_{L+R}$
in the Higgs sector\cite{cus}.
The deviation of $\tan\beta$ from unity measures the symmetry
breaking, which is  induced  at the loop level due to
the Yukawa interactions.
Thus, the enhanced symmetry can be  responsible for
the smallness of the region.
It also puts an upper bound on $m_{t}^{pole} \simle 185 $ GeV
so that $\tan\beta < 2$.

As mentioned above, we will not discuss all possible signatures,
but will focus on the loop-induced Higgs boson mass.
(Another interesting signature is a
possible light mixed $t$-scalar \cite{dreesetal}.)
In particular, Barger et al. argued \cite{barger},
using the leading-logarithm approximation,
that $m_{h^{0}} \simle 85$ GeV for $m_{t}^{pole} \simle 160$ GeV.
We will show that this is rather a typical mass range
and that the upper bound, though still relevant for
LEPII, is much higher. An important enhancement
comes from the large left-right mixing
in the $t$-scalar sector and from the large split between
the two $t$-scalar mass eigenstates.
That enhancement leads to the upper bound
$m_{h^{0}} \simle 100$ (110) GeV for $m_{t}^{pole} \simle 160$ (175) GeV.
The enhancement is sensitive to the type of vacuum stability
constraints one imposes (see below).

%4444444444444

{}From the minimization of the Higgs scalar potential one
has for the supersymmetric Higgsino mass parameter
$\mu^{2} \propto 1/[\tan^{2}\beta - 1]$,
and  $\mu^{2}$ diverges as $\tan\beta \rightarrow 1$,
i.e., in the symmetric limit.
(In fact, one expects finite loop
corrections to be relevant near the limit, and we
can assume, without loss of generality, $\tan\beta \simgr 1.1$
\cite{bando}.)
The large $\mu$-parameter dictates the
phenomenology of the scenario. In particular,
the tree-level CP even mass matrix is nearly degenerate:
it has a very heavy and a nearly zero mass eigenvalue
(this is a trivial consequence of the custodial symmetry).
In practice, the light tree-level eigenvalue
$m_{h^{0}}^{T}< M_{Z}|\cos 2\beta|$ is small
but grows  with $m_{t}^{pole}$ (i.e., with the lower bound on $\tan\beta$).
At the one-loop level, one has $m_{h^{0}}^{2}
= {m_{h^{0}}^{T}}^{2} + \Delta_{h^{0}}^{2} \approx \Delta_{h^{0}}^{2}$.
The upper bound on the loop correction can be estimated\cite{ln}
(assuming only $t$-scalar contributions):
\begin{equation}
m_{h^{0}}^{2} \leq M_{Z}^{2}\cos^{2}2\beta
 + \frac{3\alpha m_{t}^{4}}{4\pi s^{2} (1 -s^{2})M_{Z}^{2}}
\left\{\ln\left(\frac{m_{\tilde{t}_{1}}^{2}m_{\tilde{t}_{2}}^{2}}
{m_{t}^{4}}\right) + \Delta_{\theta_{t}} \right\}
\label{mh}
\end{equation}
where
\begin{eqnarray}
&\Delta_{\theta_{t}} =
\left(m_{\tilde{t}_{1}}^{2}-m_{\tilde{t}_{2}}^{2}\right)
\frac{\sin^{2}2\theta_{t}}{2m_{t}^{2}}\ln\left(
\frac{m_{\tilde{t}_{1}}^{2}}{m_{\tilde{t}_{2}}^{2}}\right)
& \nonumber \\
&  + \left(m_{\tilde{t}_{1}}^{2}-m_{\tilde{t}_{2}}^{2}\right)^{2}
\left(\frac{\sin^{2}2\theta_{t}}{4m_{t}^{2}}\right)^{2}
\left[2 - \frac{m_{\tilde{t}_{1}}^{2}+m_{\tilde{t}_{2}}^{2}}
{m_{\tilde{t}_{1}}^{2}-m_{\tilde{t}_{2}}^{2}}
\ln\left(\frac{m_{\tilde{t}_{1}}^{2}}{m_{\tilde{t}_{2}}^{2}}\right)
\right]. &
\label{mixing}
\end{eqnarray}
Due to the large $\mu$ parameter the left-right mixing $\theta_{t}$
between the two $t$-scalars can be substantial
and the leading-logarithm and the mixing $\Delta_{\theta_{t}}$ terms
in (\ref{mh}) can be equivalent. Thus,
$m_{h^{0}}$ is enhanced
by an additional factor of $\sim \sqrt{2}$.
However, since
$\tan\beta$ increases with $m_{t}$ (from perturbativity),
$|\mu|$ and $\Delta_{\theta_{t}}$ decrease and
we obtain an interesting interplay between the overall
factor of $m_{t}^{4}$ and $\Delta_{\theta_{t}}$.

%555555555555555555

Eq. (\ref{mh}) is, of course, only an approximation.
The results of a complete calculation for the Higgs boson mass
are shown in Fig. 3
(for the details of the calculation, see Ref. 4).
$\tan\beta$ is constrained as in Fig. 1,
we assume the gaussian
distribution $m_{t}^{pole} = 143 \pm 18$ GeV
(from precision data\footnote{Using the recent CDF range\cite{cdf}
for $m_{t}^{pole}$ would only change the relative
population in the different histograms
and would not affect our discussion.})
and all superpartner masses are constrained to be $\simle 1$ TeV.
The monte-carlo generated
histograms contain information on the
upper bounds and on the distributions.
The upper bound is a function of $m_{t}$:
\begin{itemize}
\item $m_{h^{0}} \simle 100$ GeV for $m_{t}^{pole} \simle 160$ GeV,
\item $m_{h^{0}} \simle 110$ GeV for $m_{t}^{pole} \simle 175$ GeV,
\item $m_{h^{0}} \simle 120$ GeV for $m_{t}^{pole} \simle 185$ GeV.
\end{itemize}
The distribution
has two peaks, one which is enhanced by the $t$-scalar mixing term and which
determines the upper bound, and a peak at a much lighter mass
from points with no enhancement.
Thus, the mass is typically much lighter than the upper bound,
and, if the $t$-quark is lighter than $\sim 170$ GeV,
may still be relevant for LEPI.
(Note that we do not attempt to give rigorous bounds.)

We compare the above bounds to those derived
for any $\tan\beta$ in Fig. 4. The $\sim 30$ GeV difference
is due to the non-vanishing tree-level mass $\simle M_{Z}$
in the general case.
(Loop corrections are, however, typically smaller in the general case.)
One can also
compare either case with typical upper bounds
$m_{h^{0}} \simle 130 - 150$ GeV derived from perturbativity
considerations \cite{upperbound}, and to the
$\underline{\mbox{lower}}$ bound given
by Sher \cite{sher} for the non-supersymmetric case,
$m_{h^{0}} \simgr 132$ GeV. However, strong assumptions
regarding vacuum stability are made in the latter case.

Complications in the calculations, which are
explored elsewhere \cite{penn594}, suggest even stronger upper bounds:

1. The magnitude of the two-loop terms is known to be small
and probably negative\cite{twoloop}.
However, the exact magnitude is ambiguous given the
complicated low-energy structure of the model.
For example, In Fig. 3  we used the effective potential
method\cite{epm} to extract $m_{h^{0}}$.
In that method $\Delta_{h^{0}}$ is given to a
one-loop leading-logarithm order.
It is straightforward
to show that the renormalization-point dependence of
the one-loop leading logarithm [e.g., in (\ref{mh})]
is equivalent to
two-loop next-to-leading logarithms
which are typically positive.
Thus, one could  overestimate $m_{h}^{0}$ when choosing
the renormalization point $Q = M_{Z}$ (as in Fig. 3).
In Fig. 5a $Q = 600$ GeV, and indeed lower masses are suggested.
In particular, if the scalar quarks are of the order of TeV or heavier,
one would falsely get large Higgs boson masses using that method,
unless, $Q$ is properly
adjusted (so that the loop expansion does not break down).
Alternatively, we could use different methods, e.g.,
the renormalization group method \cite{rgm} leads to typically
lower one-loop upper bounds (in agreement with Fig. 5a).
This is because
one is implicitly including negative two-loop
leading logarithm contributions in that method.
In short, Fig. 3 corresponds to a conservative
estimate of the upper bounds.
Once the non-trivial task
of a complete two-loop calculation
is carried out, we expect slightly stronger upper bounds.

2. In some cases large $t$-scalar mixing corresponds to the
SM model populating only a local minimum of the full scalar potential.
That is, when considering all scalar fields on equal footing,
the global minimum may be in a direction in which
the $t$-scalar has a vacuum expectation value and, thus, the
physical vacuum is not stable.
Eliminating
cases in which a global color and charge breaking minimum
exists slightly reduces the upper bound
and also affects the mixing peak population.
The effect is reduced with increasing $m_{t}$ and
never diminishes the mixing peak completely.
One could try to identify dangerous directions in the multidimensional
scalar-field space and characterize those directions
by analytic constraints.
However, such constraints entail assumptions
that are not always useful for identifying the
deepest minimum (i.e., the constraints are too weak). In addition,
analytic constraints typically do not distinguish
a local minimum from a global one
and can exclude legitimate points (i.e., are too strong).
Thus, analytic treatments are often not satisfactory.
Furthermore,
a point in parameter space which is consistent with
electroweak symmetry breaking
contains at least a local minimum
which does not respect color and charge symmetries.
This is shown in Ref. 4 where we also
compare numerical and analytic treatments of the
problem. Recent progress was also reported by Carena and Wagner
\cite{ccb2}.

Lastly, the prediction of $m_{h^{0}}$ is relatively
stable when considering model-dependent evolution of
the soft parameters between the Planck and GUT scales \cite{new}.
This is shown in Fig. 5 and is due to the cancellations between
corrections
to the relevant soft parameters and to $\mu^{2}$.
(However, the $t$-scalar mixing and mass are affected.)
Thus, our discussion is only slightly affected by
that uncertainty.

%777777777777777777777777

To conclude,
we studied the one-loop Higgs boson mass in
the region of parameter space  which is well
motivated by Yukawa unification.
We treat all
theoretical uncertainties in a reasonable manner and
the upper bounds given, e.g.,
$m_{h^{0}}^{2} \simle 110$ GeV
(for $m_{t}^{pole} \simle 175$ GeV), correspond to
the most conservative estimate.
The Higgs boson in this scenario is most likely in a range
that is, in principle, visible
to LEPII [unless the $t$-quark  is heavier than $\sim 180$ GeV,
in which case there is not much motivation to consider region ($i$)].
Two-loop calculations and vacuum stability constraints
slightly strengthen the upper bounds.

\subsection*{Acknowledgments}
This work was done with Paul Langacker
and in part with Alex Pomarol, and
was supported by the US Department of Energy
Grant No. DE-AC02-76-ERO-3071.

%FFFFFFFFFFFFFFFFF
\newpage

\begin{figure}
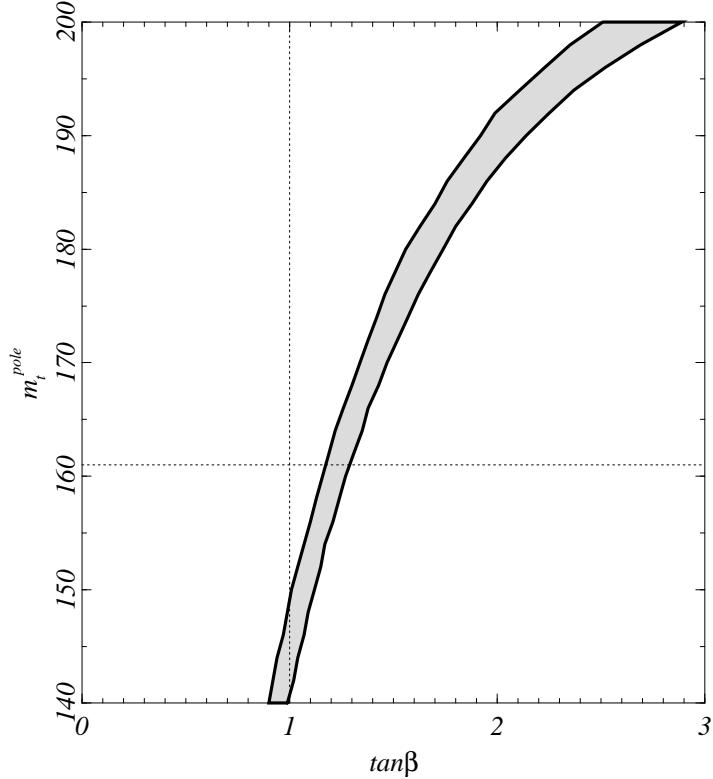

\caption{The allowed region for $\tan\beta \approx 1$.
The $t$-quark pole mass $m_{t}^{pole}$ is in GeV.
The lower bound on $\tan\beta$ is from $h_{t} < 3$
at the GUT scale, and
the upper bound from bottom-tau unification
(including theoretical uncertainties). The one s.d. range
for $m_{t}^{pole}$, suggested by precision data, and $\tan\beta = 1$
are indicated for comparison.
Recent CDF analysis suggests $m_{t}^{pole} \approx 174 \pm 16$ GeV.
For $\tan\beta < 1$ no model with universal boundary conditions
is possible.}
\end{figure}

\begin{figure}
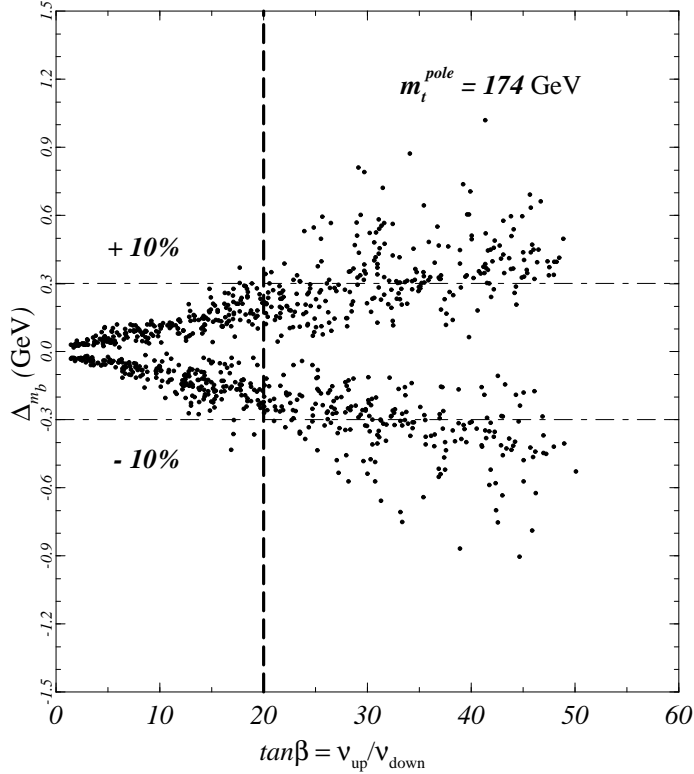

\caption{
The correction to the $b$-quark
$\overline{\mbox{MS}}$ mass $m_{b}(M_{Z}) \approx 3$ GeV (in GeV)
from finite superpartner-loops.
$m_{t}^{pole} = 174$ GeV and universal boundary conditions
for the soft parameters at the unification scale are choosen
at random. All points included are consistent with
electroweak breaking. $\tan\beta = 20$ is indicated for comparison.
For $\tan\beta < 20$ corrections are less than $10\%$.}
\end{figure}

\begin{figure}
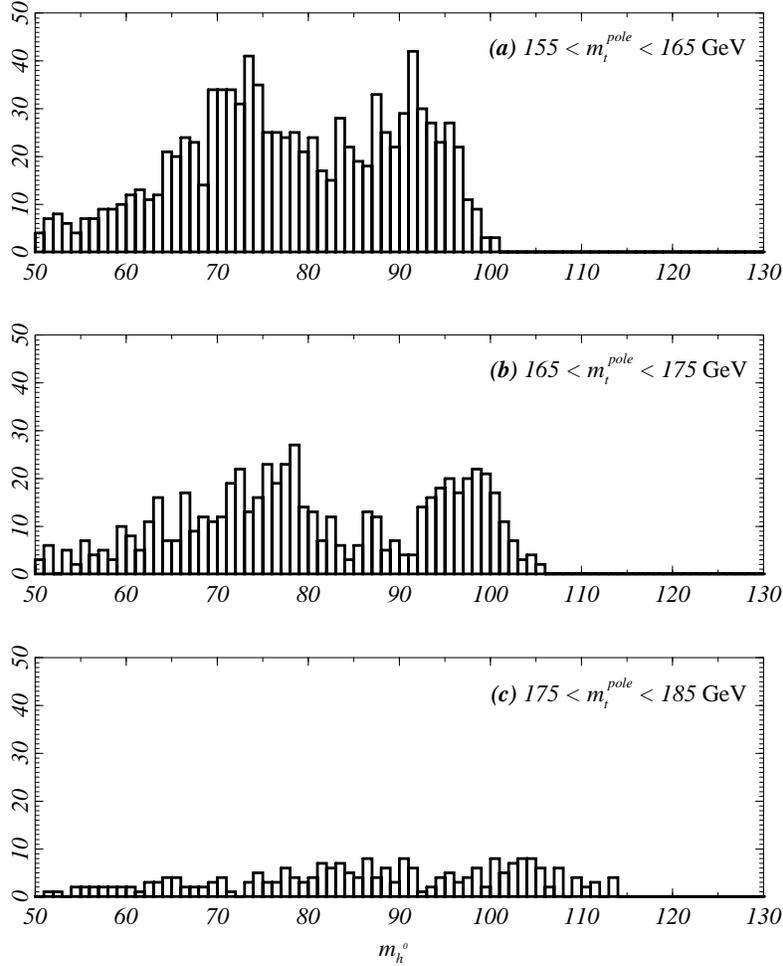

\caption{
The Higgs boson mass $m_{h^{0}}$ (in GeV) distribution in a sample of
monte-carlo calculations with random universal boundary conditions at
the unification scale and using the effective potential method
with subtraction scale $M_{Z}$. $\tan\beta$ is constrained
as in Fig. 1. $m_{t}^{pole}$ is in the range ($a$) [155, 165],
($b$) [165, 175], ($c$) [175, 185] GeV and has the gaussian
distribution $143 \pm 18$ GeV.}
\end{figure}

\begin{figure}
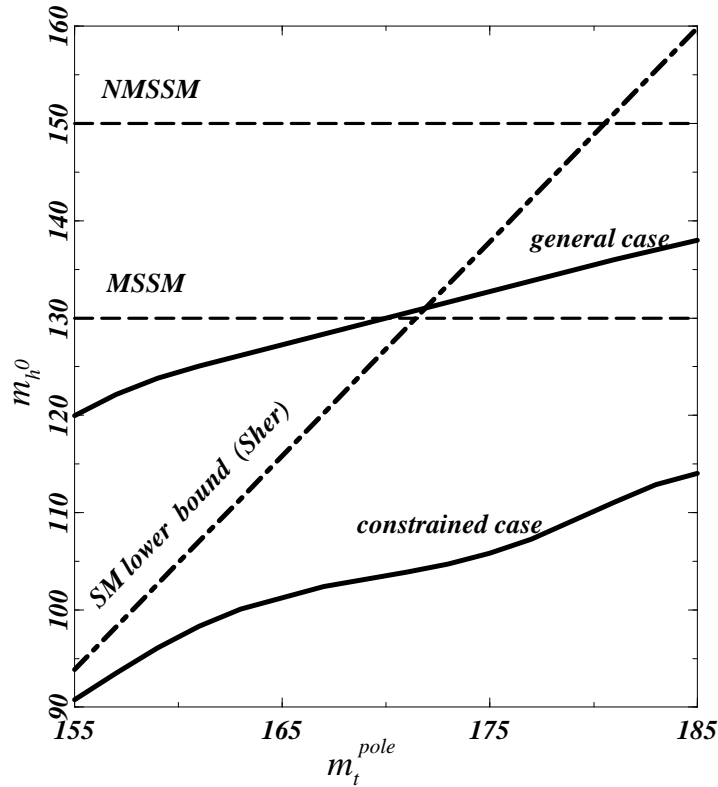

\caption{
The upper bound on the Higgs boson ($\overline{\mbox{MS}}$)
mass $m_{h^{0}}$
as a function of $m_{t}^{pole}$ (all masses are in GeV).
In the constrained case $\tan\beta$ is constrained by
bottom-tau unification as in Fig. 1. The general case
is for all $\tan\beta$ (consistent with electroweak breaking).
Note the local minimum in the constrained curve which is due
to the interplay between the mixing enhancement and
the $m_{t}^{4}$ factor.
The upper bound curves are derived using monte-carlo methods
and are not rigorous. Generic upper bounds in the MSSM
and in non-minimal models (NMSSM) derived from perturbativity
considerations are shown for comparison.
Also shown a suggestive lower bound in the non supersymmetric case
(we use $\alpha_{s} = 0.125$ in the formula of Ref. 17).}
\end{figure}

\begin{figure}
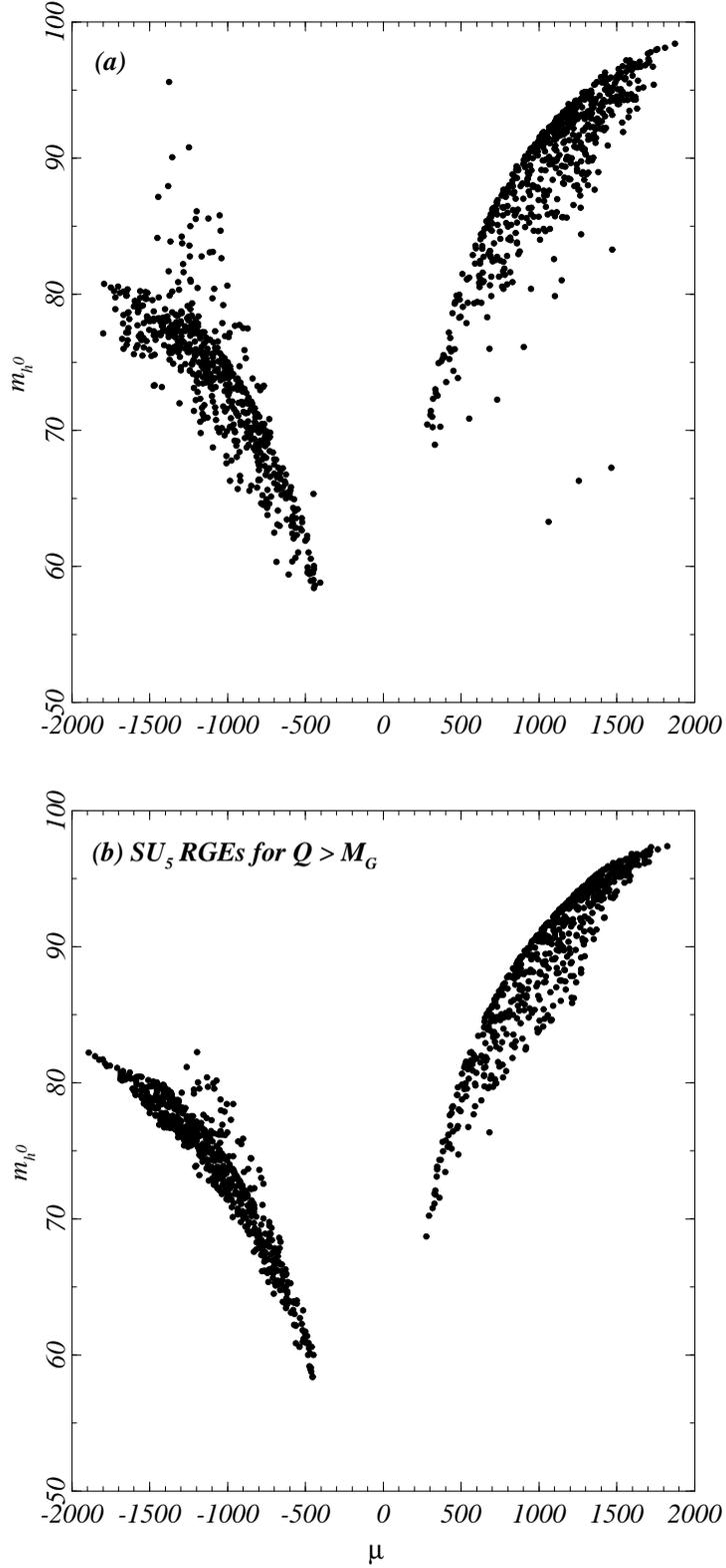

\caption{
The Higgs boson ($m_{h^{0}}$) $vs.$ Higgsino ($\mu$) mass plane
(in GeV) for
$m_{t}^{pole} = 174$ GeV, $\tan\beta = 1.52$, and with
random universal boundary conditions at ($a$)  the GUT scale
$\approx 10^{16}$ GeV, ($b$) at the (reduced) Planck scale
$\approx 10^{18}$ GeV. In case ($b$)
we assume a minimal $SU_{5}$
model above the GUT scale (for details, see Ref. 22).
The effective potential method with a subtraction scale $Q = 600$ GeV is
employed, which leads to lower values of $m_{h^{0}}$ than for $Q = M_{Z}$.
}
\end{figure}

\end{document}